# Development of a neural network to recognize standards and features from 3D CAD models

Alexander Neb[a,*], Iyed Briki[a]; Raoul Schoenhof[a]

[a]*Fraunhofer Institute of Production Engineering and Automation, Nobelstr. 12, 70569 Stuttgart, Germany*

* Corresponding author. Tel.: +49 711 970 1353; fax: +49 711 970 1008. E-mail address: alexander.neb@ipa.fraunhofer.de

**Abstract**

Focus of this work is to recognize standards and further features directly from 3D CAD models. For this reason, a neural network was trained to recognize nine classes of machine elements. After the system identified a part as a standard, like a hexagon head screw after the DIN EN ISO 8676, it accesses the geometrical information of the CAD system via the Application Programming Interface (API). In the API, the system searches for necessary information to describe the part appropriately. Based on this information standardized parts can be recognized in detail and supplemented with further information.

## 1. Introduction

Process planning in assembly is a time-consuming and costly process which often still requires a lot of human effort. Teams of experts are working intensively on analyzing 3D CAD data sets and technical drawings to obtain information for process planning in assembly. Nevertheless, errors can occur which may cause long delays. In addition, this process is carried out several times for different products, also for small batch sizes, due to the rapidly changing market and customer requirements.

An optimization of process planning in assembly consists in processing the associated activities with the aid of computer-aided solutions. This avoids errors and significantly reduces the processing time. In addition, this also involves the automation of process planning in assembly. A tool for generating the optimal assembly sequence with regard to cost-effectiveness and adherence to schedules is required. As part of a Fraunhofer IPA project, this is being worked on. An expert system [1] processes information describing the assemblies of the product and uses it to generate an optimal assembly sequence.

An existing SolidWorks API tool [2] provides the input data for the expert system. So, the tool automatically determines the spatial relationships between the parts in an assembly. These relationships describe the types of contacts which exist between the parts. In addition, the masses, volumes and bounding boxes of the parts can also be extracted from the data set of their CAD models. The information content of these data is extended in the context of this work.

The existing expert system [1,2] is already capable of automatically generating assembly sequences from a 3D CAD model of an assembly. However, the automatic identification of parts is still an unresolved issue. This identification is needed for an automatic determination of the required assembly tools and defining factors like tool changing costs or generation of assembly criteria like the connection type.

Till this day, designers struggle in defining standardized parts like bolts, nuts and retaining rings. This is especially a problem if the designers try to reuse the CAD data for downstream processes such as assembly planning. It is the case that parts are defined down to the smallest detail by their standard, but still cannot be recognized or reused. With the help of this additional information, assembly sequences can be determined even more precisely and efficiently. In addition, the automatic recognition of norms and standards provides a completely new way of analyzing 3D models with a view to optimize them. For this special purpose, a system was developed which extracts every single part of a CAD model of an assembly, converts and analyses it with the aid of a neural network pertaining the parts' standards.

## 2. State of the art

The aim of this work is to recognize the standardized parts in an assembly model of a CAD system. A classification tool can solve this task. This requires, in a first step, a descriptor to model the parts. There are two types of shape descriptors: view-based descriptors and 3D shape descriptors [3].

The view-based descriptor model objects are based on their views or projections. Wu and Jen [4] developed a method to classify prismatic parts based on their three projections in x, y, and z directions. These are represented by rectilinear polygons, skeletons and then vectors. The vectors are used as an input for a neural network to classify them. Unfortunately, the developed tool can only be applied to prismatic parts.

Other view-based descriptors are classified into contour based and region based approaches. These are the Fourier descriptors and the Zernike moments descriptors [5]. The Fourier descriptors are coefficients of Fourier transformations of the function which describes the distance between contour and centroid. By contrast, the Zernike moment descriptors only analyze the interior of the shape [5]. These descriptors can be combined into one classifier. Chen et al. [6] use it for the light field descriptors. This classifier focuses on analyzing a group of views of an object from different perspectives. Qin et al. [7] prove that Zernike's moment descriptors outperform the Fourier descriptors and the combination of both descriptors. However, the classes used for this comparison have a wide range and just classify different part types in the same class.

Su et al. [3] developed a view-based classifier for 3D objects which reaches an accuracy of 90.1%. The input of this tool is a set of pictures, which is processed as an array of pixels. By means of Convolutional Neural Networks (CNNs), the considered object can be classified.

The mentioned approaches use pictures or views of objects to classify them. By contrast, there are also approaches which are based on the 3D geometries. The profile of an object can be considered as a set of form features, histograms, a point cloud or a voxel model. To solve this mentioned problem, statement of recognizing standards of these approaches are specified more in detail in four sub-approaches.

For the first approach to recognize standardized parts, a set of requirements was defined. This set refers to the characteristic form features of each class. For this, the form features of the part are recognized and compared to the requirements. The recognition of form features can be graph-based, rule-based, syntactic-pattern-based, etc. [8].

For the second approach, histograms are built to compare the geometry of parts. These histograms describe distances between randomly defined points and their frequencies [9,10] or the curvatures and their frequencies [11] on a mesh-model.

A developed tool for the third approach is PointNet [12]. PointNet classifies objects and segments of parts and scenes. The input of this tool is a matrix which describes the coordinates of each point in the point cloud. This approach leads to very promising results because also in this way, the same object with different matrices can be modeled.

For the forth approach, objects are represented with voxel models. Wu and Jen [4] classify objects by means of deep belief neural networks. The input of their tool is a voxel model with occlusion describing this object. For 10 classes, this tool reaches an accuracy of 83.54%. Another voxel-model-based tool of Qi et al. [13] processes objects in different orientations. Although the resolution of voxel models was only 30x30x30 voxels, this tool delivers results which are comparable to the multi-view CNNs [13,3].

For this reason, this work focuses on the fourth and most promising approach of voxel models to solve the problem statement of recognizing standards.

## 3. Concept framework

To achieve the main goal of recognizing standardized parts in an assembly of a CAD model also the assembly relevant features have to be extracted. These features include, after all, the geometrical properties of the standardized parts. Furthermore, for the identified screws and nuts, also the required torque needs to be readout from the CAD model.

For this purpose, a feature extraction tool was developed, by accessing the SolidWorks API. Firstly, this tool "disassembles" the assembly model and saves the parts as STEP files and SLDPRT files (SolidWorks native file format). Secondly, the tool calls a further tool "STEP-NPY-Converter" [14]. In addition, the tool converts the CAD models of the parts into numerical 3D arrays. So, the STEP files are converted into mesh models (.stl) by the python-library aoc-xchange [15]. These models are rasterized into a binary voxel grid by using, the 3D mesh voxelizer of Min [16] and Nooruddin [17] . Thus, each part in the 3D CAD model of the assembly is represented by a 3D array. The voxels referring to a free space are expressed by 0, whereas voxels indicating a sub volume of the part are denoted by 1.

The third step begins after every single CAD part is converted to voxel models. The systems then starts to call a second converting tool for the classification task. Therefore, an artificial neural network was trained, which classifies the parts into 10 classes: "hexagon head screws", "hexagon socket head cap screws", "hexagon socket countersunk head screws", "hexagon nuts", "retaining rings for shafts", "retaining rings for bores", "parallel keys form A", "parallel keys form C" and "chamfered plain washers" (table 1). The classes represent standardized parts which are very common in industry. A special challenge is the differentiation of the "parallel keys form A" and "parallel keys form C" based on their very similar shape and nearly identical form features. Thus, the capability and performance of the developed system will be judged in manner to differentiate between very similar looking standards. For each part the classification tool creates 21 invariants. These describe the same part, but in different orientations as voxel models. All invariants of each parts are classified. It should be noted that the higher the number of invariants, the higher the computation time becomes. The number of 21 invariants is thus a compromise between computation time and an efficient classifier. The output of this tool is a matrix with 21 columns. The predicted class for each invariant is given in the corresponding column. The system calculates the predictions and identifies the best solution for each part. Afterwards, only the class with the highest prediction will be considered in further steps.

Lastly, the final system reads the single SLDPRT files of the recognized standardized parts in, in order to extract the part proprieties. For this step, the SolidWorks native file format is necessary, because the voxel model itself does not contain enough geometrical information to recognize features like threads or pitch angles. For this reason, only SLDPRT files are considered for the task. Additionally, in this step, the results of classification should be verified in order to extract the right information for the recognized standardized part. As a neural network cannot guarantee results with a 100% accuracy, the purpose of the validation task is to detect the failures within the classifying. Some requirements for each class can be defined and checked. These requirements consider the geometry and the number of shell surfaces on parts of each class. If the defined requirements are satisfied, the developed system begins with the extraction of the class proprieties. This task is mainly based on the analysis of shell surfaces of the parts, the distance between them and the defined features in the modeling process. The developed system recognizes the following standards: DIN EN ISO 4017, DIN EN ISO 4014 and DIN EN ISO 8676 for hexagon head screws; DIN EN ISO 4762 for hexagon socket head cap screws; DIN EN ISO 10642 for hexagon socket countersunk head screws; DIN EN ISO 4032, DIN EN ISO 8673, DIN EN ISO 4033, DIN EN ISO 8674, DIN EN ISO 4035 and DIN EN ISO 8675 for hexagon nuts; DIN 471 for retaining rings for shafts; DIN 472 for retaining rings for bores; DIN 6885 for parallel keys; and DIN EN ISO 7090 for chamfered plain washers.

## 4. Recognition of standards and features

### 4.1. Classification

To identify the described standards, nine classes of standardized parts in an assembly were defined. Those have to be recognized by the system (Table 1). To realize this classification capability, a neural network was trained and its performance was tested. In order to be successful, this step requires the generation of datasets in beforehand.

**Datasets:** CAD models, especially for standardized parts, are available in libraries of common CAD software. Mainly, for this work, the SolidWorks toolbox library and the database of the Fraunhofer IPA was used. It facilitates the automatic generation of various CAD models. These are considered as raw data. The number of the generated models for each class is given in Table 1.

At first, each version of the standardized parts is represented in the raw data only once. This means that the considered parts are only available in one orientation. Nevertheless, the designer can choose an arbitrary direction while modelling parts. The developed system is able to recognize the standardized parts, independent of its orientation. To guarantee this, for each version of the parts in the raw data, several invariants with different orientations for data augmentation are generated.

Table 1. Raw data of the classification.

| Class | Class name | No. of models |
|---|---|---|
| 1 | hexagon head screws | 173 |
| 2 | hexagon socket head cap screws | 108 |
| 3 | hexagon socket countersunk head screws | 314 |
| 4 | hexagon nuts | 46 |
| 5 | retaining rings for shafts | 24 |
| 6 | retaining rings for bores | 24 |
| 7 | parallel keys form A | 189 |
| 8 | parallel keys form C | 342 |
| 9 | chamfered plain washers | 11 |
| 10 | Others | 666 |
| Sum | | 1897 |

For the classes from 1 to 9, 100.000 different invariants for each class are generated. For the 10th class, even 400.000 invariants are created, based on the random shapes. Thus, at the end, a dataset of 1.3 million models is obtained. This information is split into two datasets. The first set of data is used for training and the second one, consisting of ~55.000 models, is used for testing the trained neural network.

**Training of the neural network:** The classification of parts requires a conversion of their CAD models into a model which can be treated with a neural network. In this study, the CAD models are converted in voxel models with three different resolutions: 32x32x32 voxels, 64x64x64 voxels and 128x128x128 voxels. The higher the resolution of the voxel model is, the more form features of parts can be recognized. However, the augmentation of the resolution also increases the computing effort.

Table 2. Configurations of the classification and the networks

| Config. | Resolution | Layers | Transfer function | Learning rate |
|---|---|---|---|---|
| 1 | $32^3$ | 8CNN+3FCL | ReLU | $5 \times 10^{-6}$ |
| 2 | $64^3$ | 8CNN+3FCL | ReLU | $5 \times 10^{-6}$ |
| 3 | $128^3$ | 8CNN+3FCL | Leaky_RelU | $5 \times 10^{-6}$ |
| 4 | $128^3$ | 8CNN+3FCL | Leaky_RelU | $5 \times 10^{-5}$ |
| 5 | $128^3$ | 8CNN+3FCL | Leaky_RelU | $1 \times 10^{-4}$ |
| 6 | $128^3$ | 6CNN+3FCL | Leaky_RelU | $5 \times 10^{-5}$ |

However, also other factors of the training process which are analyzed in this study. These are the number of layers, the transfer functions and the learning rates, which are presented in Table 2. The used network consists of eight CNN layers and three Fully Connected Layers (FCL). On the sixth configuration, the number of layers was reduced to six. To develop a classifier for voxel models with the resolutions of $32^3$ and $64^3$ voxels, the transfer function Rectified Linear Unit (ReLU) is employed. While training the network for models with a higher resolution, the results were not satisfactory. Thus, the leaky ReLU replaces the previous transfer function.

To optimize the weights and biases of the network while training, the ADAM-Algorithm [18] is applied. For this gradient-based algorithm, three different learning rates are analyzed: $6 \times 10^{-6}$, $6 \times 10^{-5}$ and $1 \times 10^{-4}$. This algorithm minimizes the following cost function for the evaluation of the deviation between nominal and actual outputs of the neural network. At this, $x_{soll,i}$ and $x_{ist,i}$ are the inputs of the nominal and actual output vectors of a model:

$$f_{cost} = \sum (x_{soll,i} - x_{ist,i})^4. \tag{1}$$

To train the network of the first configuration, a number of 250.000 epochs are executed. In every epoch, 32 models are treated as one batch. For the second configuration, a number of 350.000 epochs were executed. The training batches consist of 16 models. For other configurations, each batch consists of only two models, because of the limited computation capacities. For these configurations, 850.000 training epochs are executed. While training three epochs for each batch are executed.

**Testing of the Networks:** Now that all configurations have been defined in Table 1, Figure 1 shows the results of the accuracy of configuration 1 and 2 (see table 2) during the testing process with a resolution of $32^3$ voxels (orange graph) and $64^3$ voxels (blue graph). As shown in this diagram, overfitting did not occur while training. The accuracy of the first configuration converges to 90%. The augmentation of resolution from $32^3$ voxels to $64^3$ voxels increases the accuracy to 95%. For this reason, the resolution is raised up to the maximum of $128^3$ voxels.

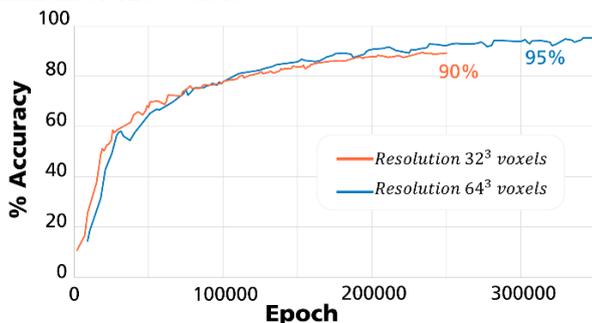

Fig. 1. Accuracy function of configuration 1 with a resolution of $32^3$ voxels and configuration 2 with a resolution of $64^3$ voxels (both with 8CNN+3FCL)

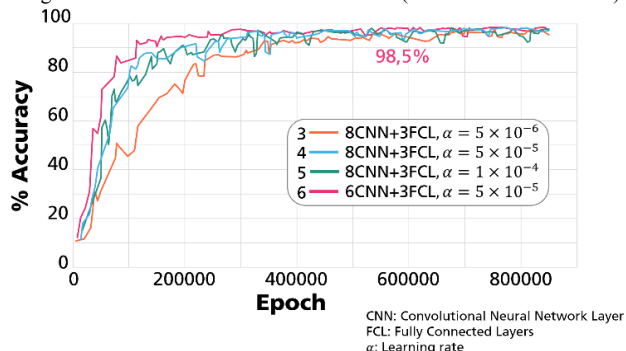

Fig. 2. Accuracy function of the third to sixth configuration with a resolution of $128^3$ voxels and different transfer functions

The results of the testing of the developed classifier for models, with a resolution of $128^3$ voxels, are shown in Figure 2. At the beginning of the training process, the accuracies continuously rise and then converge to values larger than 95%. The first configuration to converge is the sixth configuration. This configuration has the least number of layers. This reveals that the corresponding network has the smallest number of variables to be optimized during training. Thus, this configuration requires less computation time then the others to reach an optimal state. On the contrary, the accuracy of the third configuration requires a higher number of epochs to reach an optimal learning state. A reason for this delay is the smallest learning rate. Between the results of the configurations four and five, there was no significant difference being identified.

The highest accuracy to be reached was 98.5%. This was achieved by the resolution of $128^3$ voxels by the sixth configuration after 546.000 trainings epochs. Thus, a network of this configuration is used for the classification of the parts.

### 4.2. Extraction of properties

After identifying the standardized parts among the assembly, the developed system extracts its properties. For this task, the system analyses the corresponding models of standardized part as native file formats in SolidWorks. As shown in chapter 4.1, the classifier reaches an accuracy of 98.5%. The remaining error of the classifier can be compensated through processing the parts in 21 orientations. In spite of the latter, there is always an uncertainty margin. The extraction of information from the CAD models depends on the results of classification. Each class of parts has its own algorithm for data extraction. With this margin of uncertainty, the tool may crash or report errors in case of wrong classification. To solve this problem, a module for validation of the classifier results was integrated.

The validation is a rule-based procedure. For each class of parts, requirements are defined. If these requirements are approved, the validation module confirms the classification results. The properties of standardized parts are described in detail and are fixed in standards. Thus, requirements are defined which are valid for all parts of the class. These requirements are related to the geometry of the parts. In general, four types of requirements are specified.

The first type concerns the number of shell faces that build the body of the parts. The geometry of these faces is analyzed within the second type of requirements. By means of the SolidWorks API, through the ISurface interface, it can be determined if the treated face is a plane, a cylinder, a cone, a sphere, etc. However, only the three first geometry types are considered. These types of geometries almost represent all geometries of the treated classes of standardized parts.

In the third type of requirements, the number of edges of faces is determined. For instance, the bottom face of a "hexagon socket head cap screw" or a "hexagon socket countersunk head screw" has seven edges, from which one is the outer bound. The other six are the edges of the "hexagon socket".

Within the fourth type of requirements, the dimensions of the CAD models with the dimensions defined on the standards were compared. This countercheck, however, is not applied for all classes. This is based on the fact that the standard specification for some standardized parts (e.g. "retaining rings for shafts") the geometry of the body is variable specified. Therefore, the first three levels cannot be applied.

To validate the classification results requirements from these different types are combined. For example, the CAD model of a "hexagon socket head cap screw" should only contain two cylindrical faces, just two or three conical faces and two plane faces: one of them only has one edge and the others have seven edges.

After classifying the parts and validating the delivered results, the developed tool extracts the properties of the recognized standardized parts. For the identification of the treats of screws and nuts, several additional values, like key width, diameter, thread pitch, the thread length, height, strength class and the assembly torque are required.

To identify the key width, it is necessary to recognize six plane surfaces with the same area in the CAD model. Here, the key is the largest width with the closest distance between faces of these six planes. This distance is determined through the function ModelDoc2::ClosestDistance in the SolidWorks API. In order to determine the height, the tool identifies faces with a distance between them corresponding to height. For instance, the height of a screw is the closest distance between two plane surfaces which only have seven edges. To extract the properties of the thread, the tool looks for a feature in the feature manager, with the type "CosmeticThread". In this feature, all required information can be found. The designer of the CAD model defines the strength class in naming of the configuration. The developed tool analyzes this string and finds out which strength class is chosen. The assembly torque is a characteristic for parts with threads. Usually, designers calculate this torque and fill it in the Bill-of-Material (BOM) table. The tool searches, based on that process, for a feature with the type "BomFeat" to access to the BOM table, where the required information can be found.

For "retaining rings for shafts", "retaining rings for bores" and "chamfered plain washers" the thickness and the suitable shaft or bores diameter are needed. The thickness is the closest distance between the largest plane faces. The suitable shaft diameter for "retaining rings for shafts" is determined by means of the cylindrical surface with the fourth largest radius. For washers, the cylindrical surface with the smallest radius should be identified. To find the suitable bore diameter for "retaining rings for bores", the cylindrical surface with the largest diameter is recognized.

Properties of parallel keys are the height, the width and the length. The height is the closest distance between the largest plane surfaces. The width is the closest distance between the next two planes. The length is the closest distance between the cylindrical surfaces with the largest radius plus the width.

After the identification of all these properties, the developed system is suitable to define standards and its features for the nine defined machine elements (elementary parts of a machine, like screws or nuts) directly in a 3D model of a CAD system.

## 5. Evaluation

**Case study 1:** The CAD model of the first case study is represented in figure 3 and represents a modified version of the cranfield assembly benchmark [19]. This model contains standardized and non-standardized parts.

This assembly consists of two "baseplates" (2). In-between these baseplates is a "separator" (1), a "shaft" (7) and two "conus bolts" (6). The "shaft" guides the "pendulum" (4), which is screwed with a "head" (5). To join the "baseplates", two "hexagon socket head cap screws" (3), two "hexagon nuts" (8) and two "chamfered plain washers" (9) are used. Both screws and nuts have a metric regular thread M12. According to the corresponding standards, the key width is 10 mm for the screws and 18 mm for the nuts. The length of the screws is 80 mm and the length of the thread is 36 mm. The height of the nuts is 10.8 mm. The required torque for the fixation of the screws and nuts is set to 1 Nm in the BOM-able. Their strength class is also set in the naming of the configuration as 10.9. The corresponding shaft diameter for the used washers is 12 mm and their thickness is 2.5 mm.

The developed system has to recognize three couples of parts as standards, namely 3, 8 and 9 and needs to classify them due to their corresponding classes.

The developed system classifies the parts in an error-free way. Only the screws, nuts and washers are classified in the corresponding classes. All other parts are set on the class of non-standardized parts (tenth class). Furthermore, the system also determines the features of the standardized parts correctly with the described properties in chapter 4. The corresponding standard designations are also delivered to the user. These are ISO 4762 - M12x80 - 10.9, ISO 4032 - M12 -10.9 and ISO 7090 – 12.

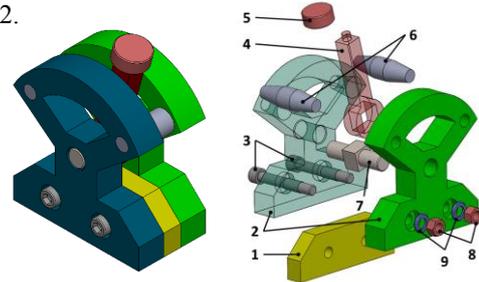

Fig. 1. Modified case study of the cranfield assembly benchmark

**Case study 2:** The CAD model of this case study is shown in figure 4 and it represents a spur gear unit. The parts of this assembly which should be identified as standardized parts are four "hexagon socket head cap screws" (1), five "retaining rings for shafts" (2) and a "parallel keys form A" (7). Admittedly, the "roller bearing" (3) and the "gear wheels" (5) are also standardized parts, however they are not considered in this work. Furthermore, the assembly also contains two "shafts" (4), two "housings parts" (6) and a "separator" (8) as non-standardized parts.

The screws have a metric regular thread M5 with a length of 22 mm. The whole length of each screw is 30 mm. According to the standard, the key width is 4 mm. The strength class is set to 8.8. In the BOM-table the assembly torque is set to 2 Nm. In this assembly there are four "retaining rings for shafts" that fit on "shaft" with 10 mm diameter. The fifth one fits on a "shaft" with 15 mm diameter. All of them have a thickness of 1 mm. The used "parallel keys" are different and can be differentiated adequately. One of them has the dimensions 5x5x12 mm and the other one 3x3x11 mm. The performance of the developed tool is also validated in this case study. These standard designations are determined as ISO 4762 – M5x30 – 8.8, DIN 471 – 10x1, DIN 471 – 15x1, DIN 6885 – A – 5x5x12 and DIN 6885 – A – 3x3x11. The results of the classification and the extraction of properties are also error-free for this case study.

In both case studies, all standardized parts were recognized correctly. Furthermore, also the necessary features and the properties were identified without mistakes. This underlines the functionality of the system and its precision.

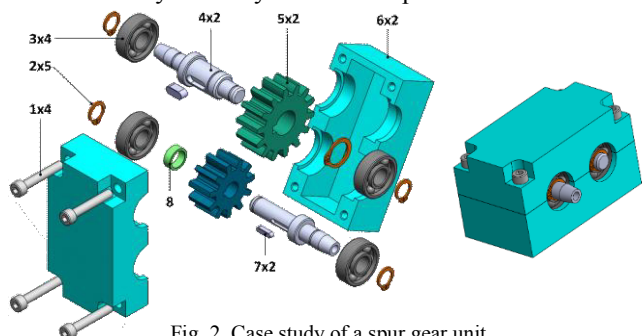

Fig. 2. Case study of a spur gear unit

## 6. Conclusion and future prospect

In this work, a novel approach was presented to recognize standardized parts directly in a 3D model of a CAD system. This procedure has the advantage of precising assembly information which is already available in the planning phase of the assembly. Many companies do not use SolidWorks tools, so standardized parts cannot always be detected by SolidWorks. The novelty of the presented tool is that it always works independently of all tools and therefore does not depend on SolidWorks. With the help of this approach it is now possible to pass on further assembly information to the planning phases in a machine-readable form.

Furthermore, the standards and features found in this work support the finding of an ideal assembly sequence significantly. This results in the fact that the expert system of Fechter und Neb [1] can be extended with these important information. Now the expert system can be extended with data on the required assembly resources such as screwdrivers or retaining rings pliers. Finally, cost functions, such as tool change costs or stability factors in assembly, can be determined.

However, a limitation of this work is the restricted amount of investigated standards. Although nine very common machine elements with several different standards were recognized, much more are relevant in an average assembly process. Especially the second case study showed that even those simple products contain more standardized parts than the ten investigated classes in Table 1. Therefore, additional machine elements will be integrated in the near future.

Furthermore, this information about standards and their related resources will also be integrated in an augmented reality assembly application [20]. The focus of this application is a guidance of less-skilled workers even through complicated assembly processes. The derived information in this work will be used to generate animations to visualize the assembly processes.

## Acknowledgements

The research presented in this paper partially received funding by the Future Work Lab Pilot Project. The Future Work Lab Pilot Project is funded by the German Federal Ministry of Education and Research (BMBF) within the Program "Innovations for Tomorrow's Production, Services, and Work" - »Future of Work« and controlled implemented by the Project Management Agency Karlsruhe (PTKA).